\documentstyle[12pt]{article}
\newcommand{\Sl}[1]{{}}           % do not show labels

\newcommand{\Section}[1]{\section{#1}\setcounter{equation}{0}}

\newcommand{\beq}[1]{\Sl{#1}\begin{equation}\if#1\empty\else\label{#1}\fi}
\newcommand{\eeq}{\end{equation}}
\newcommand{\beqa}[1]{\Sl{#1}\begin{eqnarray}\if#1\empty\else\label{#1}\fi}
\newcommand{\eeqa}{\end{eqnarray}}

\newcommand{\bm}[1]{\mbox{\boldmath$ #1 $}}
\newcommand{\bmc}{{\bm c}}
\newcommand{\bmq}{{\bm q}}
\newcommand{\bmr}{{\bm r}}

\newcommand{\vr}{{\bmr}}
\newcommand{\vq}{{\bmq}}

\newcommand{\eq}[1]{(\ref{#1})}

\newcommand{\calG}{{\cal G}}

\begin{document}

\thispagestyle{empty}

\begin{center}

{ \sl
\flushleft THU-94-12 \\
comp-gas/9501001\\[10mm]
}
{\Large{\bf
Algebraic Spatial Correlations and\\[3mm]
Non-Gibbsian Equilibrium States\footnote{
contribution to Princeton meeting `Discrete Models for Fluid Dynamics,' held
June 27-29, 1994}
}}\\[12mm]
{\large  M.H. Ernst and H.J. Bussemaker}\\
\vspace{1.2cm}
{\it Institute for Theoretical Physics\\
University of Utrecht\\
Princetonplein 5, P.O.Box 80.006\\
3508 TA Utrecht, The Netherlands}
\end{center}
\vspace{1.2cm}
\begin{abstract}
\noindent
Non-Gibbsian stationary states occur in dissipative non-equilibrium systems.
They are closely connected with the lack of detailed balance and the absence
of a fluctuation-dissipation theorem. These states exhibit spatial
correlations that are long ranged under generic conditions, even in systems
with short range interactions, provided the system has slow modes and
some degree of spatial anisotropy.
In this paper we present a theory for static pair correlations in lattice
gas automata violating detailed balance, and we show that the spatially
uniform non-Gibbsian equilibrium state exhibits long range correlations,
even in the absence of an external driving field.\\[4mm]
\end{abstract}

\newpage

\Section{Introduction}

As highlighted in a recent review by Dorfman et al.~\cite{DKS}, a major
theme in non-equilibrium statistical mechanics during the past few decades
has been the question under what conditions the correlations in fluids,
that consist of molecules with short-range interactions, can become
long-ranged.
The existence of generic long range spatial correlations in non-equilibrium
stationary
states of condensed matter (e.g. systems with a constant temperature gradient,
driven diffusive systems \cite{Lebow}) is by
now  well understood, and is intimately
connected with the existence of long time tails in Green-Kubo type time
correlation functions.
In this paper we discuss the occurrence of such correlations
in lattice gas automata (LGA's).

The long range correlations can be described at three different levels of
microscopic detail:
(i) The macroscopic phenomenological equation
(diffusion, Navier-Stokes) with a Langevin noise term added
\cite{Grin1,Grin2}.
Here the transport coefficients and noise strengths are phenomenological
input in the theory. A summary of this approach will be given in section~2.
(ii) The kinetic (Boltzmann) equation to which a Langevin noise term is added
\cite{EC-JSP}. Here the transport
coefficients are determined by the theory, but the noise strength is
phenomenological input. This intermediate level of description will not
be pursued here. For applications to LGA's we refer
to Ref.~\cite{DE-Waterloo}.
(iii) A fully microscopic statistical mechanical description. Here both the
noise strength and the transport coefficient are calculated from the theory.

To obtain a description at the fully microscopic level, one starts from the
equations of motion for the dynamic variables or the phase space density,
and then derives a BBGKY-hierarchy for the distribution functions.
After appropriate approximations one obtains
kinetic equations for the single particle and pair correlation functions.
Long range spatial and temporal correlations are created by sequences of
ring collisions.

If a classical fluid is in thermodynamic equilibrium,
long range spatial correlations
only occur under very special conditions, when parameters like the temperature
are tuned to a critical value.
In general however the correlation length is finite.
The situation is different when external driving fields or constraints imposed
by external reservoirs prevent the fluid from achieving a thermodynamic
equilibrium state satisfing the constraints of detailed balance.
In this case {\em generic} long range spatial correlations exist
in the non-equilibrium stationary state, for almost all parameter values.

The same concepts are also directly relevant for LGA's that violate the
constraints of (semi)-detailed balance. Recently, a microscopic theory at the
level of pair correlations (ring kinetic theory) has been applied to LGA's to
calculate the equal-time position and velocity correlations in spatially
uniform equilibrium states~\cite{BED-JSP}.
The primary goal of the present paper is to apply the ring kinetic theory to
demonstrate the existence of generic long range spatial correlations in uniform
equilibria of LGA's without detailed balance.
This will be worked out in section 3. The coarsest description, by means of a
fluctuationg diffusion equation, is conceptually the simplest, but it captures
all essential features. It will be discussed in section 2.

All three levels of description share the same essential ingredients for the
existence of such algebraic correlations: (i) the existence of local
conservation laws or slow (diffusive or hydrodynamic) modes; (ii) lack
of detailed balance; (iii) some degree of anisotropy, either due to the
underlying lattice, or due to an external driving field or gradient imposed
by external reservoirs.

\Section{Langevin Equation}

A clear presentation of the basic theory of non-equilibrium Langevin
models has been given by Grinstein et al. \cite{Grin1,Grin2}. Here we only
indicate the essential steps. Consider a locally
conserved density $h(\vr ,t)=\langle \hat h(\vr,t) \rangle$.
On average it satisfies a diffusion equation
with an anisotropic diffusion tensor. To
account for the fluctuations, a Langevin noise term is added to this equation.
It reads after Fourier transformation,
\beq{a1}
\partial_t \hat{h}(\vq , t) = (D_x q^{2}_{x} + D_y q^{2}_{y} ) \hat{h} (\vq
,t ) + \hat{\eta} (\vq, t),
\eeq
where the Gaussian random noise is determined by
\beq{a2}
\langle \hat{\eta}(\vq ,t) \hat{\eta}(-\vq,t^\prime ) \rangle =
2(B_x q^{2}_{x} + B_y q^{2}_{y} ) \delta (t-t^\prime ).
\eeq
We allow for spatial anisotropy by having two constants $B_x$
and $B_y$. The diffusion coefficients $D_\alpha$ and noise strengths $B_\alpha$
are phenomenological coefficients; they are not provided by the theory.
Without loss of generality we use
a two-dimensional presentation to denote longitudinal and transverse
directions with respect to the wave vector $\vq$.

The equal time density-density correlation function
in the stationary state, $G(\vr , \infty ) \equiv {\cal G}(\vr )$, is given by
\beq{a3}
{\cal G}(\vr ) = \int \frac{d\vq}{(2\pi )^d}
e^{{\textstyle i}\vq\cdot \vr} \chi (\vq ).
\eeq
Here $d$ is the number of dimensions and the stationary susceptibility can be
calculated from \eq{a1} and \eq{a2} as,
\beq{a4}
\chi (\vq ) = \lim_{t\rightarrow\infty} \langle | \hat{h}(\vq ,t) |^2 \rangle
= \frac{B(\vq )}{D(\vq )} = \frac{B_x \hat{q}^{2}_{x} + B_y \hat{q}^{2}_{y}}
{D_x \hat{q}^{2}_{x} + D_y \hat{q}^{2}_{y}},
\eeq
where $\hat{q}_\alpha = q_\alpha / | \vq |$ is a Cartesian component of a
unit vector $(\alpha = x,y, \ldots , d)$.\\[5mm]
\noindent
{\em Possible scenario's:}
\begin{itemize}
\item[(i)] If the system satisfies {\em detailed balance}, i.e. if
the equilibrium distribution is given by the Gibbs distribution, the
equilibrium
value of the susceptibility $\chi (\vq ) \simeq \chi_0$ as $q \to 0$
is known from thermodynamics, and transport coefficients and noise strengths
are related
by the {\em fluctuation-dissipation} theorem, $B(\vq ) = \chi_0 D(\vq )$.
The correlation function is short-ranged, $G(\vr ) \simeq \chi_0 \delta (\vr
)$.
\item[(ii)] However {\em without} the detailed balance constraint, the
equilibrium state is not a thermodynamic state, and the corresponding
susceptibilities are unknown. There is {\em no} fluctuation-dissipation
theorem imposing a relationship between $B(\vq )$ and $D(\vq )$. The long
wavelength limit $\chi_0 (\hat{q})$ of the susceptibility is in general
{\em anisotropic} and depends on the direction of $\hat{q}$. A rescaling
of the integration variables in (\ref{a3}), namely $\vq = {\bm k}/r$,
shows that $G(\vr ) \sim r^{-d}$ at large distances. As diffusive
modes are correlated over time intervals $t \sim r^2$, the spatial
correlations $\sim r^{-d}$ have an intimate connection with the
long time tails $t^{-d/2}$ in the velocity and other current-current
correlation functions \cite{EuroP}. For later reference
we quote the explicit form of the correlation function (\ref{a3}) at large
distances in two dimensions, where the susceptibility is given by (\ref{a4}),
i.e.\
\beq{a5}
G(\vr ) = G(x,y) \simeq \frac{1}{\pi \sqrt{D_x D_y}}
\left( \frac{D_x B_y - D_y B_x }{D_y x^2 + D_x y^2} \right).
\eeq
As we shall see in section 3, this scenario is realized in LGA's where the
collision rules do not satisfy detailed balance {\em and} break the
lattice symmetry by having different transition probabilities in $x$- and
$y$-direction.
\item[(iii)] Different scenario's are realized when the phenomenological
$B(\vq )$ and/or $D(\vq )$, and therefore also $\chi(\vq)$ in (\ref{a4}),
have a weaker anisotropy as $q \to 0$, that only shows up in terms of
relative order ${\cal O}(q^2)$ or ${\cal O}(q^4)$. Then $G(\vr )$ decays
again algebraically on account of \eq{a3}, with a tail proportional to
$1/r^{d+2}$ or $1/r^{d+4}$ respectively.
\end{itemize}

\noindent
Suppose that one wants to describe a system where the microscopic transition
probabilities have the symmetry of a $d$-dimensional lattice, but where
there is {\em no} externally imposed direction in which the {\em lattice
symmetry is broken}. Examples are the spin-spin correlations on hypercubic
lattices as discussed by Grinstein \cite{Grin2}, or several non-detailed
balance LGA's on square, triangular, or FCHC lattices (see sections 3 and 4).

The general form of the noise strength for small $q$ is
\beq{a6}
q^2 B(\vq ) = q_\alpha q_\beta B_{\alpha\beta} + q_\alpha q_\beta q_\gamma
q_\delta B^4_{\alpha\beta\gamma\delta} + {\cal O}(q^6),
\eeq
and a similar expansion holds for the transport coefficient $q^2 D(\vq )$.
Greek
indices denote Cartesian components and summation convention is used
for repeated indices. The coefficients $B_{\alpha\beta}, D_{\alpha\beta}$ and
$B^4_{\alpha\beta\gamma\delta}, D^4_{\alpha\beta\gamma\delta}$ are
respectively tensors of rank 2 and rank 4. On all lattices with 4- or 6-fold
symmetry,
the second rank tensors are {\em isotropic}, i.e. $B_{\alpha\beta} =
B\delta_{\alpha\beta}$.

Fourth rank tensors $\bm{B}^4$ and $\bm{D}^4$ are only {\em isotropic} on the
two-dimensional triangular lattice and the four-dimensional FCHC lattice, and
{\em anisotropic} on all other lattices in two and three dimensions. In the
latter
cases the susceptibility for $q \to 0$ behaves as $\chi (\vq ) =
\chi_0 + q^2 \chi_2 (\hat{q})$ and $G(\vr ) \sim 1/r^{d+2}$. In the former
cases the susceptibility has the form $\chi(\vq ) = \chi_0 + q^2 \chi_2 +
q^4 \chi_4 (\hat{q})$ and $G(\vr ) \sim 1/r^{d+4}$. For later reference
we calculate the results for a square lattice. There $\chi_2 (\hat{q}) = E_s +
E_a \left(\hat{q}^{4}_{x} + \hat{q}^{4}_{y}\right)$, and $G(\vr )$ in
\eq{a3} approaches
\beq{a7}
G(\vr ) \simeq - 6E_a / \pi r^4 \qquad (r \to \infty ) .
\eeq
In  the microscopic LGA's violating detailed balance, to be discussed in
section 3 and 4, the coefficients $B(\vq )$ and $D(\vq )$ can be calculated
explicitly from the ring kinetic theory, and yield the coefficients of the
algebraic tails in $G(\vr )$.

\Section{Ring Equation for LGA's}

In this section we present the ring kinetic theory for velocity and spatial
correlations in LGA's
violating detailed balance with respect to the Gibbs distribution.
Such correlations have been observed in several models by various authors
\cite{Dubrulle,Henon,BE-Nice}.

Starting from a BBGKY hierarchy for the $n$-particle distribution
functions, we use cluster expansion techniques to derive approximate
kinetic equations.
In zeroth approximation the standard nonlinear Boltzmann equation is obtained;
the next approximation yields the ring kinetic equation, similar to that for
hard sphere systems, describing the time evolution of pair correlations. The
derivation of the generalized Boltzmann equation and ring equation can be
found
in Ref.~\cite{BED-JSP}. Here we only sketch the results in so far as they are
needed for the present analysis.
Similar cluster expansion techniques have been developed in
Ref.~\cite{Boghosian}.

The collision step of a LGA is defined by the strictly local
 transition probabilities
$A_{s\sigma}$ from a precollision state $s$ at a single node
to a postcollision state $\sigma$ at the same node. It is followed
by a propagation step.

Much is known about LGA's that satisfy the condition of
 semi-detailed balance, $\sum_s A_{s\sigma} = 1$,
or (stronger) detailed balance, $A_{s\sigma} = A_{\sigma s}$, with respect
to the Gibbs distribution. The equilibrium Gibbs
distribution depends only on the globally conserved quantities, such
as the total number of particles $N$, the total momentum $P$ or the total
energy $E$. The Gibbs distribution is completely factorized: no
equal-time correlations between occupation numbers exist.

In LGA's without detailed balance $\sum_s A_{s\sigma} \neq 1$.
Starting from a completely factorized precollision state,
the collisions will then create postcollision correlations between
occupation numbers at the same node.
The  postcollision on-node correlation for $i \neq j$ are given by
\cite{BED-JSP,BE-Nice}
\beq{b1}
    \langle \sigma_i \sigma_j \rangle^*
	\equiv \sum_{s\sigma} \sigma_i \sigma_j A_{s\sigma} F(s) \neq f_i f_j ,
\eeq
where we have introduced the average occupation number
$f_i = \langle s_i \rangle$.
The completely factorized distribution function $F(s)$ is given by
\beq{b2}
	F(s) = \prod_i f_i^{s_i} (1-f_i)^{1-s_i}.
\eeq

Due to subsequent propagation and collision steps the postcollision
on-node correlations will be  transformed into on- and off-node
precollision correlations, via a process of scattering
by other particles in the system (ring collisions).

If we neglect all correlations between occupation numbers, the time
evolution of the one-particle distribution function,
$f_i({\bf r},t) = \langle s_i(\bmr,t) \rangle$, is given by the
 nonlinear Boltzmann equation \cite{DE-Waterloo},
\beq{b3}
	f_i(\bmr+\bmc_i,t+1) = f_i(\bmr,t) + \Omega_i (f(\bmr,t)),
\eeq
where $\vr$ labels the nodes and $\bmc_i$ the nearest neighbor links.
For the detailed definition of $\Omega_i (f)$ in terms of $A_{s\sigma}$ we
refer to Ref.~\cite{BED-JSP}.
At a more refined level of description we include all pair correlations
between fluctuations ($\delta s_i = s_i - f_i$),
\beq{b5}
	G_{ij}(\bmr, \bmr', t)
		\equiv \langle \delta s_i(\bmr,t) \delta s_j(\bmr',t) \rangle.
\eeq
The coupled time evolution of $f_i(\bmr,t)$ and $G_{ij}(\bmr,\bmr',t)$ is
then given by the {\em generalized} Boltzmann equation
\beq{b6}
  f_i(\bmr+\bmc_i,t+1) = f_i(\bmr,t) +
	\Omega_i(f(\bmr,t)) + \sum_{k<\ell}
	\Omega_{i,k\ell}(f(\bmr,t)) G_{k\ell}(\bmr,\bmr,t),
\eeq
together with the {\em ring} kinetic equation
\beq{b7}
  G_{ij}(\bmr+\bmc_i,\bmr'+\bmc_j,t+1) =
	\delta(\bmr,\bmr') B_{ij}(\bmr,t) +
	\sum_{k,\ell}\omega_{ij,k\ell} G_{k\ell}(\bmr,\bmr',t).
\eeq
Here the linear pair collision operator is given by
\beq{b8}
  \omega_{ij,k\ell} =
	\{\delta_{ik}+\Omega_{ik}(f(\bmr,t))\}
	\{\delta_{j\ell}+\Omega_{j\ell}(f(\bmr,t))\}.
\eeq
In the equations above we have introduced $\Omega_{ik} = \partial\Omega_i /
\partial f_k$ and $\Omega_{i,k\ell} = \partial^2 \Omega_i / \partial f_k
\partial f_\ell$.
The on-node source term $B_{ij}(\bmr,t)$ depends on $f_j(\bmr,t)$ and
$G_{ij}(\bmr,\bmr,t)$, and essentially represents the correlations created
by the collision step.

The explicit form of the matrix elements $\Omega$ are not needed in the
present
analysis, where we focus on the large-$\vr$ behavior of the solution of
\eq{b7} in the uniform equilibrium state. The on-node source term contains a
dominant contribution, referred to as the simple ring approximation, which
will be given below for the special case of equilibrium. Here we neglect
subleading terms,
linear in the pair correlation function \cite{BED-JSP}, which correspond to
the  `repeated ring approximation' in the theory of continuous fluids.

A spatially homogeneous equilibrium solution obeys the relation
$f_i(\bmr,\infty)$ $= f_i$ and
$G_{ij}(\bmr,\bmr',\infty) = \calG_{ij}(\bmr-\bmr')$.
For a given average occupation $f_i$, the correlations can be obtained from
a linear equation,
\beq{b9}
	\calG_{ij}(\bmr) = v_0 \int_{\rm 1BZ} \frac{{\rm d}{\bmq}}{(2\pi)^d}
		e^{{\textstyle i} {\bmq}\cdot\bmr} \left[
		\frac{1}{1-s(\bmq)\omega} s(\bmq) B \right]_{ij},
\eeq
with $v_0$ the volume of the unit cell of the direct lattice
($v_0=1$ for square and cubic lattices and $v_0=\frac{1}{2}\sqrt{3}$
for triangular lattices), and the pair streaming operator given by
\beq{b10}
	s_{ij}(\bmq)=\exp[i\bmq\cdot(\bmc_j - \bmc_i)].
\eeq
The on-node source term
\beq{b11}
	B_{ij} 	= B_{ij}(\bmr,\infty)
		= \langle \sigma_i \sigma_j \rangle^* - \langle s_i s_j
\rangle
\eeq
equals the change in on-node correlations caused by the collision step in
equilibrium, where $\langle \ldots \rangle^*$ is defined in \eq{b1}
($B_{ij}=0$ in SDB-models).

The spatial correlation functions of most interest are those between conserved
densities, defined as
\beq{b12}
\calG_{ab}(\vr) = \sum_{ij} a_i b_i \calG_{ij}(\vr)
\equiv \langle ab | \calG(\vr) \rangle ,
\eeq
where $a_i$ and $b_i$ are collisional invariants. In a purely diffusive
system $a_i = b_i = 1$, in a thermal fluid-type LGA $a_i , b_i
= \{ 1, c_{\alpha i},{ 1\over 2 }c^{2}_{i} \}$. Moreover, we note the
important conservation law $\langle ab | B \rangle = 0$ as implied by
(\ref{b11}), stating that the source term $B$ is orthogonal to products of
collisional invariants.

The numerical calculation of $\calG_{ij}(\bf 0)$ for given $f_i$ involves
matrix inversion and integration over the first Brillouin zone.
Since $f_i$ again depends on $\calG_{ij}(\bf 0)$, we use an iterative
scheme to find a self-consistent solution $\{f_i,\calG_{ij}(\bf 0)\}$.
Once $\calG_{ij}(\bf 0)$ is known, all off-node correlations
$\calG_{ij}(\bmr)$ can be calculated from it in a straightforward
way. For details we refer to Ref.~\cite{BED-JSP}.

The analytical calculation of \eq{b9} can be carried through for large
spatial distances $(r \to \infty )$ or small wave numbers $(q \to 0)$.
The essential observation is that the {\em pair}
operator $s(\bmq)\omega$ in \eq{b9} has {\em slow} (diffusive) modes
$\chi_{\mu\nu}(\vq )$ with eigenvalues behaving like $\Lambda_{\mu\nu}
(\bmq)
\simeq 1-q^2 D_{\mu\nu}(\hat{q}) + \ldots $ for small $q$, where $D_{\mu\nu}
(\hat{q})$ may depend on the direction of $\vq$. The slow modes give rise
to singular denominators in \eq{b9}, and are responsible for the large-$\bmr$
behavior of $\calG_{ab}(\vr)$.

To analyze this more systematically it is convenient to make a spectral
decomposition of the pair operator $s(\vq )\omega$ in \eq{b9}, i.e.
\beq{b13}
  \calG_{ab} (\vr ) \simeq v_0 {\sum_{\mu\nu}}^* \int_{1BZ}
  \frac{{\rm d}\vq}{(2\pi )^d}
  e^{{\textstyle i}\vq \cdot\vr} \frac{\langle ab | \chi_{\mu\nu}(\vq )\rangle
  \langle \tilde\chi_{\mu\nu}
  (\vq ) s(\vq ) | B \rangle}{1- \Lambda_{\mu\nu}(\vq )}.
\eeq
Here $\chi_{\mu\nu}$ and $\tilde\chi_{\mu\nu}$ are respectively right and
left eigenfunctions (product modes) of the pair operator $s(\bmq)\omega$ with
eigenvalue $\Lambda_{\mu\nu} (\vq )$. These product modes form a complete
bi-orthogonal set of basis functions in the pair space, i.e.\
$\langle \tilde{\chi}_{\mu\nu} | \chi_{\mu^\prime \nu^\prime}\rangle =
\delta_{\mu\mu^\prime} \delta_{\nu\nu^\prime}$. The asterisk in \eq{b13}
indicates that the $\mu\nu$-summation is restricted to pairs of slow
(hydrodynamic or diffusive) modes.

To make these results more explicit we specialize to a purely diffusive model
on a
square lattice. Here $a_i = b_i =1$ are the only collisional invariants, and
the only
slow pair mode $\chi_{DD}(\vq )$ is a product of single particle diffusive
modes. The conservation laws imply $\langle 11 | B\rangle = 0$, and the lattice
symmetry guarantee that
$\langle \tilde{\chi}_{DD}(\bmq) s(\bmq) | B \rangle = q^2
B(\vq )$ where $B(\vq )$ has the general form \eq{a6}. The
denominator, $1-\Lambda_{DD}(\vq ) = 2q^2 D(\vq )$, has a similar form
for small $q$.  Moreover, one can show that $\langle 11 | \chi_{DD}\rangle =
1 + {\cal O}(q^2 )$. The (non)-isotropy of the second and fourth rank
tensors $\bm{B},\bm{D}$ and $\bm{B}^4,\bm{D}^4$ respectively, depends on
the microscopic collision rules of the lattice model under consideration.

In {\em summary}, we have established that the spatial correlation
functions $\calG_{ab}(\vr )$ in equilibrium approach for large $r$ the generic
form \eq{a3}
and \eq{a4} of the phenomenological theory of section 2, and in addition
we have provided expressions
for the tensors $\bm{B},\bm{D}$ and $\bm{B}^4,\bm{D}^4$, leading to a
complete microscopic theory for long range behavior of the spatial correlation
functions of LGA's violating the detailed balance constraints.

\Section{Interacting Random Walkers}
\subsection{Spatial Symmetry}

To illustrate the general results of the previous sections we first
consider a model of interacting random walkers on a square
lattice with Fermi exclusion.
There is no breaking of the lattice symmetry in the microscopic transition
probabilities. The model allows for at most {\em one} particle per state
$\{ \vr , \bmc_i \}$ ($i$ = 1,2,3,4) where $\vr$ labels a lattice
node and $\bmc_i$ a nearest neighbor link. The in-state or pre-transition
state of the node is described by the set of occupation numbers
$s(\vr ) = \{ s_i (\vr ); i = 1,2,3,4 \}$, and similarly the out-state
or post-transition state is described by
$\sigma (\vr ) = \{ \sigma_i (\vr ); i = 1,2,3,4 \}$.
The dynamics consists of local transitions, followed by propagation. The
local transitions conserve the number of particles.

If there is only {\em one} particle  at node $\vr $, it behaves like a
random walker; if there are {\em three} particles at $\vr $, or equivalently
one hole,
the hole behaves like a random walker. If there are {\em no} particles or no
holes, in- and out-states are identical. To describe the {\em two}-particle
interactions, let $\{ ij \}$ denote the two-particle state in which links
$i$ and $j$ are occupied. Then the transitions are defined as,
\beq{c1}
\{ 13 \}, \{24 \}
%\stackrel{\stackrel{\textstyle\alpha}{\rightarrow}}
%{\stackrel{\textstyle\leftarrow}{\beta}}
\begin{array}{c}
\alpha \\ \vspace{-3mm} \longrightarrow \\ \longleftarrow \\ \beta
\end{array}
\{ 12 \}, \{ 23 \}, \{34 \}, \{ 41 \}
\eeq
where $\alpha$ denotes the transition probability of any state on the left
to any on the right, and $\beta$ the probability for the reverse
transition. We note that the states on the right, and those on
the left of \eq{c1} are not connected by lattice symmetries. If we
denote the transition probabilities among the four states on the right hand
side
by $\gamma$, and those among the ones on the left by $\delta$, then we have
the normalization conditions
$4\alpha + 2 \delta = 1$ and $2\beta + 4\gamma = 1$.

In case $\alpha \neq \beta$ the matrix of transition rates,
$A_{s\sigma} \neq A_{\sigma s}$, does not obey the (semi)-detailed balance
conditions of  section 3. The transitions in \eq{c1}
create correlations between different velocity links, and the
pre-transition correlations $\langle s_i s_j \rangle$ differ from the
post-transition ones $\langle \sigma_i \sigma_j \rangle^*$ in \eq{b11},
even in the uniform equilibrium state. Consequently, $B_{ij}$ in
\eq{b11} is nonvanishing, and so is $q^2 B(\vq )$ in \eq{a6}. Because
of the square symmetry the second rank tensor
$B_{\alpha\beta} = B\delta_{\alpha\beta}$
in \eq{a6} is isotropic, but the fourth rank tensor
$B^4_{\alpha\beta\gamma\delta}$ is not, which leads immediately
to the $1/r^4$ tail in \eq{a7}. The coefficient $E_a$
is proportional to $(\alpha - \beta ) f^2 (1-f)^2$ where
$f = {1 \over 4} \rho$ is the reduced average density.

In case $\alpha = \beta $ the {\em detailed balance} constraints are
satisfied, and the transitions do {\em not} create any correlations in
the stationary state, i.e. $\langle \sigma_i \sigma_j \rangle^* = \langle
s_i s_j \rangle = f_i f_j$, and long range correlations are absent.
The details of these calculations and the explicit value of the coefficient
$E_a$ of the algebraic tail will be given elsewhere \cite{BE-long}.

\subsection{Broken Spatial Symmetry}

The long range correlations will be much stronger if the microscopic
transition probabilities are anisotropic. It is a driven diffusive system
that remains spatially uniform in the stationary state.
For purpose of illustration we
consider again a diffusive 4-bit LGA on a square lattice where the local
transition probabilities in $x$- and $y$-directions are different, i.e.
\beq{c2}
A_{s\sigma} = \frac{1}{Z(s)} \delta (\rho (s),\rho (\sigma )) \exp [bj_x (s)
j_x (\sigma )] .
\eeq
Here $j_x (s) = \sum_i c_{xi} s_i (r)$
represent the local particle current in the
$\hat{x}$-direction. The local number of particles $\rho (s) = \sum_i s_i (r)$
in the in-state, as well as $\rho (\sigma )$ in the out-state is conserved,
and $Z(s)$ is a normalization constant
chosen such that $\sum_\sigma A_{s\sigma} = 1$.
For positive $b$-values transitions with parallel in- and out-currents in
the $x$-direction are favored; those with anti-parallel currents are
suppressed. There is no bias in the $\hat{y}$-direction.
If $b=0$ the LGA satisfies detailed balance and represents random walkers
interacting through the Fermi exclusion rule. All spatial correlations are
vanishing. If $b \neq 0$ there exists a spatially uniform equilibrium state
with long range correlations $\sim 1/r^2$. The qualitative features of this
model are similar to those of the diffusive lattice gas, studied in
\cite{EuroP}.

It is important to emphasize that the constraints of (semi)-detailed balance,
i.e.\
\beqa{c3}
\sum_s  F_0(s) A_{s\sigma} = F_0 (\sigma ) & \qquad & (SDB) \nonumber \\
        F_0(s) A_{s\sigma} = F_0 (\sigma ) A_{\sigma s} & \qquad & (DB)
\eeqa
refer to the factorized Gibbs distribution $F_0 (s) = C \exp [\alpha\rho (s)
+ \ldots ]$, which depends on the state variable $s$ only through the
conserved densities $\rho (s)$, etc. This implies $F_0 (s) = F_0 (\sigma )$
and the
$SDB$ constraints reduce to the familiar form $\sum_s A_{s\sigma} = 1$.
Similarly the $DB$ constraints in \eq{c3} reduce to
$A_{s\sigma} = A_{\sigma s}$.

Therefore the model properties $\sum_s A_{s\sigma} \neq 1$ or $A_{s\sigma}
\neq A_{\sigma s}$ do exclude the existence of the Gibbsian equilibrium
state, but do {\em not} necessarily imply the existence of position and
velocity {\em correlations} in equilibrium. This statement can be slightly
rephrased by
saying that the model properties $\sum_s A_{s\sigma} \neq 1$ or $A_{s\sigma }
\neq
A_{\sigma s}$ do {\em not exclude} the existence of a completely
{\em factorized equilibrium} distribution. We present an example.

Consider a variation on the LGA in \eq{c2} of a driven diffusive system, where
\beq{c4}
A_{s\sigma} = \frac{1}{Z(s)} \delta (\rho (s),\rho (\sigma )) \exp
[{\bm j}(\sigma )\cdot {\bm E}].
\eeq
Here $\mbox{\boldmath $E$}$ is an external bias field and
$\mbox{\boldmath $j$}(s) = \sum_i \mbox{\boldmath $c$}_i s_i (r)$
the local (non-conserved) particle current. Then
clearly, $\sum_s A_{s\sigma} \neq 1$. Nevertheless, the SDB- or DB-condition
\eq{c3} does admit a completely factorized (normalized) equilibrium
solution,
\beq{c5}
F_0 (s) = C \exp [ \alpha\rho (s) + \mbox{\boldmath $j$} (s) \cdot
\mbox{\boldmath $E$}],
\eeq
which can also be cast in the standard form \eq{b2} with an
equilibrium single particle distribution of Fermi-type,
\beq{c6}
f_i = \left[1 + \exp(-\alpha-\mbox{\boldmath $c$}_i \cdot
\mbox{\boldmath $E$} )\right]^{-1}.
\eeq
Establishing the stability of this equilibrium state \eq{c5}-\eq{c6} requires
an $H$-theorem. In fact, at this meeting  H. Chen \cite{HChen} has presented a
proof of the H-theorem for a lattice gas that satisfies the DB-condition
\eq{c3} with respect to a factorized equilibrium distribution \eq{c5} with
$\exp [ - {\bm j} (s) \cdot {\bm E}]$ replaced by
$\Pi_j (F_j / I_j)$, where $F_j$ and $I_j$ are arbitrary positive constants.

Concluding, we can say that the model satisfies the SDB-constraints
\eq{c3} with respect to the {\em factorized} distribution $F_0(s)$. The
source term $B_{ij}$ in \eq{b11} vanishes for this model and consequently
all on- and off-node correlations vanish in the equilibrium state.

\Section{Conclusion}

We have studied LGA's with local (on-node) interactions that possess slow
(diffusive or hydrodynamic) modes and lack (semi)-detailed balance (SDB)
with respect to the Gibbs distribution. As is well known
\cite{DKS,Lebow,EuroP}, for long times such systems approach non-thermodynamic
spatially uniform states, which
exhibit long range correlations, if the isotropic spatial symmetry is broken by
externally applied fields or reservoirs.
The surprising result is that there exist LGA's that exhibit long-range
spatial correlations $\sim r^{-d-2}$ in the equilibrium state without any
fields
or reservoirs that break the spatial symmetry. These correlations are absent
in LGA's satisfying (semi)-detailed balance. The property $\sum_s A_{s\sigma}
\neq 1$, i.e.\ violation of the standard SDB-constraints, does not necessarily
imply the existence of long range correlation, nor does it exclude the
existence
of a completely factorized equilibrium distribution. This was shown by
explicit construction of a diffusive LGA model that satisfies the
SDB-constraints with respect
to a non-Gibbsian factorized equilibrium distribution (see \eq{c5} and
\eq{c6}).
A similar fluid-type LGA has been presented at this meeting by
H. Chen \cite{HChen}.
In continuous fluids long-range correlations and non-detailed balance
stationary states
can only exist in open systems, externally driven or constrained by
reservoirs,
because the thermodynamic (Gibbs) equilibrium state satisfies detailed
balance.

\vspace{1cm}

\noindent
\section*{Acknowledgements}
The authors wants to thank J.R. Dorfman and J.W. Dufty for stimulating
discussions. H.J.B.\ acknowledges
financial support by the ``Stichting voor Fundamenteel Onderzoek der Materie''
(FOM), which is sponsored by the ``Nederlandse Organisatie voor
Wetenschappelijk Onderzoek'' (NWO).

\end{document}